\documentclass[footinbib,aps,pra,reprint,superscriptaddress,nopacs]{revtex4-1}

\usepackage{graphicx}
\usepackage{amsmath,amsfonts,amssymb,bbm,bm}
\usepackage{verbatim}
\usepackage{braket}
\usepackage{mathtools}
\usepackage[absolute]{textpos}
\usepackage{ragged2e}

\usepackage{xcolor}

\newcommand{\imag}{\mathrm{i}}
\def\sF{\mathcal{F}}
\def\sP{\mathcal{P}}
\def\sH{\mathcal{H}}
\DeclareMathOperator{\Tr}{Tr}

\begin{document}
\title{High-dimensional discrete Fourier transform gates with \\the quantum frequency processor}

\author{Hsuan-Hao Lu}
\affiliation{Elmore Family School of Electrical and Computer Engineering and Purdue Quantum Science and Engineering Institute, Purdue University, West Lafayette, Indiana 47907, USA}

\author{Navin B. Lingaraju}
\affiliation{Elmore Family School of Electrical and Computer Engineering and Purdue Quantum Science and Engineering Institute, Purdue University, West Lafayette, Indiana 47907, USA}
\affiliation{SRI International, Arlington, Virginia 22209, USA}

\author{Daniel E. Leaird}

\author{Andrew M. Weiner}
\affiliation{Elmore Family School of Electrical and Computer Engineering and Purdue Quantum Science and Engineering Institute, Purdue University, West Lafayette, Indiana 47907, USA}

\author{Joseph M. Lukens}
\email{lukensjm@ornl.gov}
\affiliation{Quantum Information Science Section, Oak Ridge National Laboratory, Oak Ridge, Tennessee 37831, USA}
\date{\today}

\begin{abstract}
The discrete Fourier transform (DFT) is of fundamental interest in photonic quantum information, 
yet the ability to scale it to high dimensions depends heavily on the physical encoding, with practical recipes lacking in emerging platforms such as frequency bins.
In this Letter, we show that $d$-point frequency-bin DFTs can be realized with a fixed three-component quantum frequency processor (QFP), simply by adding to the electro-optic modulation signals one radio-frequency harmonic per each incremental increase in $d$. We verify gate fidelity $\sF_W>0.9997$ and success probability $\sP_W>0.965$ up to $d=10$ in numerical simulations, and
experimentally implement the solution for $d=3$, utilizing measurements with parallel DFTs to quantify entanglement and perform full tomography of multiple two-photon frequency-bin states. Our results furnish new opportunities for high-dimensional frequency-bin protocols in quantum communications and networking. 
\end{abstract}

\maketitle
\textit{Introduction.---}The existence of incompatible observables in quantum theory represents one of the central departures of quantum from classical mechanics, underpinning the Heisenberg uncertainty principle and precluding the actualization of quantum states with, e.g., arbitrarily well-defined position and momentum. 
In finite-dimensional Hilbert spaces, incompatibility appears in the form of mutually unbiased bases (MUBs)~\cite{Wootters1989, Durt2010}. Consider two orthonormal $d$-dimensional bases $\{\ket{\alpha_m}\}$ and $\{\ket{\beta_m}\}$; they comprise a pair of MUBs if and only if they satisfy $\left|\braket{\alpha_m|\beta_{m'}}\right|^2=\frac{1}{d}$ for all $m,m' \in\{0,1,...,d-1\}$, which implies that measurements in the $\alpha$-basis provide no information about the results of measurements in $\beta$, and vice versa.

MUBs are optimal measurements for tomography of noisy quantum states~\cite{Wootters1989, Adamson2010}, expose tampering from eavesdroppers in quantum key distribution (QKD)~\cite{Scarani2009, Cerf2002, Sheridan2010}, and provide efficient entanglement witnesses~\cite{Spengler2012, Coles2017}. One archetypal pair of MUBs are the logical and discrete Fourier transform (DFT) bases: $\{\ket{m}\}$ and $\{\ket{f_m}\}$, where 
$\ket{f_m} = \frac{1}{\sqrt{d}}\sum_{n=0}^{d-1} e^{-2\pi\imag mn/d} \ket{n}$. Measurements with both have been utilized extensively in a variety of photonic quantum information experiments. In time-bin encoding, the DFT has been realized with nested delay interferometers, supporting both optical frequency division multiplexing~\cite{Hillerkuss2010, Hillerkuss2011} and high-dimensional QKD~\cite{Islam2017a, Islam2017b}; in path encoding, DFT operations 
have been used for on-chip state characterization~\cite{Wang2018a} and 
three-photon bosonic coalescence~\cite{Spagnolo2013}; and in orbital angular momentum, spatial light modulators have enabled measurements in DFT bases for tomography~\cite{Giovannini2013} and entanglement certification~\cite{Bavaresco2018, Ecker2019}. 

In frequency-bin encoding, the quantum frequency processor (QFP)---a concatenation of alternating electro-optic phase modulators (EOMs) and pulse shapers~\cite{Lukens2017}---has enabled experimental demonstration of DFT gates up to $d=3$, using a three-element QFP (two EOMs and one pulse shaper)~\cite{Lu2018a}. But although theoretical and numerical results ~\cite{Lukens2017,Lukens2020a} indicate the QFP's potential to reach even higher-dimensional DFTs with additional elements, it is unclear whether more efficient DFT constructions are possible with smaller systems, a question of practical importance toward high-dimensional quantum communications and networking protocols.

\begin{figure*}[bt!]
\centering
\includegraphics[width=\textwidth]{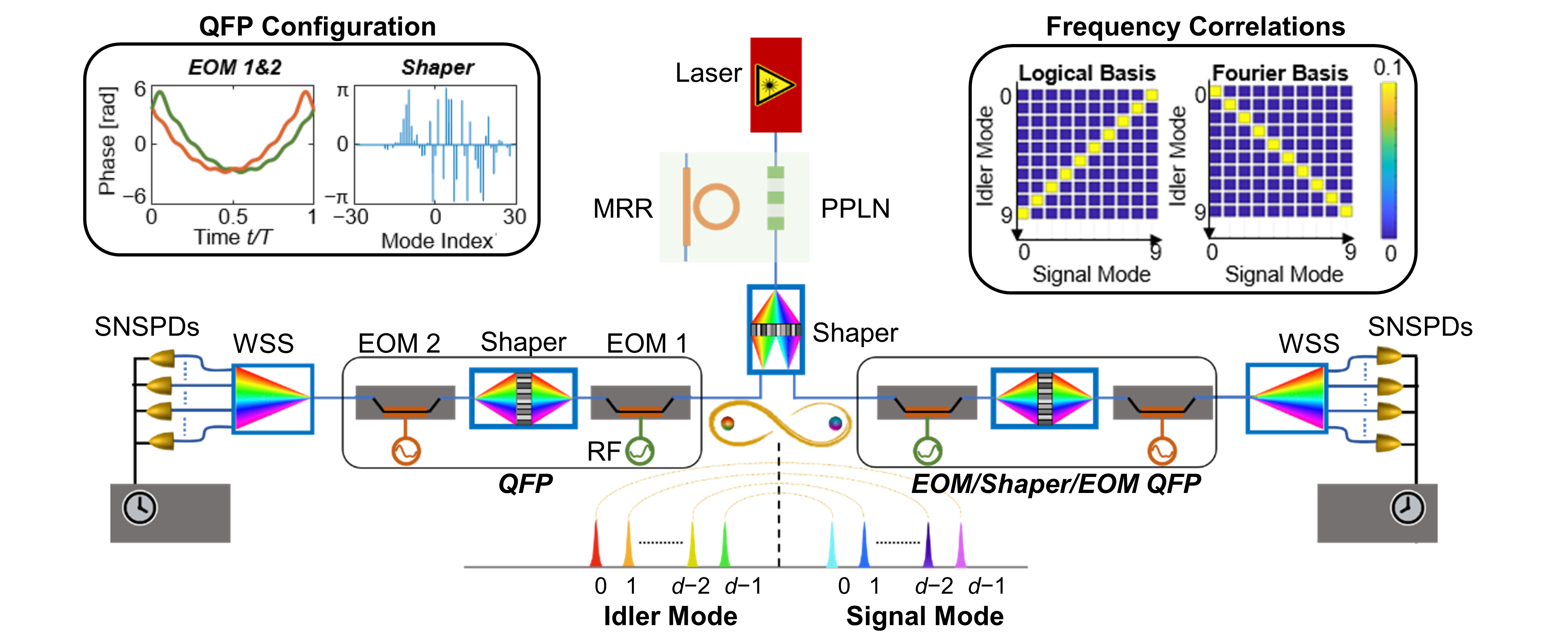}
\caption{Representative application for the frequency-bin DFT. Entangled photons are generated and sent to different users, each of whom uses a QFP to apply either the identity or $d$-dimensional DFT prior to frequency-resolved detection. Insets show simulation results for the ten-dimensional case: optimal DFT modulation patterns (upper left) and spectral correlations for an ideal maximally entangled input (upper right).
See text for details.}
\label{GenScheme}
\end{figure*}

In this work, we address this question directly and develop explicit designs for efficient frequency-bin DFTs. 
We find that a fixed-length three-component QFP is sufficient to reach DFT gate performance with fidelity $\sF_W>0.9997$ and success probability $\sP_W>0.965$ for all dimensions examined ($d\leq10$). 
The only requirement as $d$ increases is the addition of one radio-frequency (RF) harmonic per dimension increment in the EOM drive functions, so that $d-1$ total RF tones enable the $d$-point DFT with high $\sF_W$ and $\sP_W$. As examples of these designs, we experimentally implement parallel $d=3$ DFTs on multiple frequency-bin entangled states, using the measurement results to perform full state tomography and quantify entanglement through Bayesian inference. Our results provide a scalable recipe for the construction of high-$d$ frequency-bin DFTs, useful for basic communication tasks in this degree of freedom and particularly well suited to tight bin spacings envisioned in future integrated devices.

\textit{DFT gate designs.---}Figure~\ref{GenScheme} highlights an example scenario leveraging QFP-based DFT operations. A high-dimensional frequency-bin entangled state---produced, e.g., by pumping a microring resonator (MRR) or a periodically poled lithium niobate (PPLN) waveguide---is split and transmitted to two users, each of whom possess a QFP, wavelength-selective switch (WSS) and superconducting nanowire detectors (SNSPDs). By synthesizing either the identity (EOMs off) or complete DFT on the QFP, measurements of the received photon can be performed in either the logical or Fourier bases, respectively. The upper-left inset shows temporal modulation patterns and pulse shaper phases that enable a high-fidelity ($d=10$)-dimensional DFT; if the original biphoton state is maximally entangled, then joint measurement outcomes will be perfectly correlated in matched bases, as exhibited in the upper-right inset for the case of an input state with uniform phase, i.e., $\ket{\psi}=\frac{1}{\sqrt{d}}\sum_{k=0}^{d-1}\ket{k,d-1-k}_{IS}$. Correlations in such complementary bases can then be used for entanglement verification or $d$-dimensional two-basis QKD~\cite{Sheridan2010}.

Mathematically speaking, in terms of logical and DFT basis states, the goal of a DFT gate is to map $\ket{f_m}$ inputs to $\ket{m}$ outputs, which can be written in 
terms of 
the input (output) annihilation operators $\hat{a}_m$ ($\hat{b}_m$) as a matrix $\hat{b}_m = \sum_{n=0}^{d-1} (F_d)_{mn} \hat{a}_n = \frac{1}{\sqrt{d}} \sum_{n=0}^{d-1} e^{-2\pi\imag mn/d} \hat{a}_n$, where the operators apply to discrete frequency bins centered at $\omega_n = \omega_0 + n\Delta\omega$. We focus on synthesizing these gates on a three-component QFP, which collectively implements the modal transformation $W$ with elements
\begin{equation}
\label{eq:W}    
W_{mn} = \sum_{k=-\infty}^\infty d_{m-k} e^{\imag\phi_k} c_{k-n},
\end{equation}
where $m,n\in\{0,..,d-1\}$. The $c_n$ ($d_n$) coefficients are defined in the Fourier series expansions of the phase modulation transformations of the first (second) EOMs in the QFP, i.e., 
\begin{equation}
\label{eq:c}
\begin{aligned}
c_n = \frac{2\pi}{\Delta\omega} \int_{\frac{2\pi}{\Delta\omega}} dt\, e^{\imag A(t) + \imag n\Delta\omega t} \\
d_n = \frac{2\pi}{\Delta\omega} \int_{\frac{2\pi}{\Delta\omega}} dt\, e^{\imag B(t) + \imag n\Delta\omega t}, 
\end{aligned}
\end{equation}
which assumes periodicity at the bin spacing and integration over one full period. We further decompose $A(t)$ and $B(t)$---the time-dependent phases applied by the respective two phase modulators within a single QFP---as Fourier series themselves: $A(t) = \sum_{p=1}^P A_p \cos(p\Delta\omega t + \gamma_p)$ and $B(t) = \sum_{p=1}^P B_p \cos(p\Delta\omega t + \delta_p)$, where $P$ is a specified integer cutoff. The complete mapping $W$ can then be compared to the ideal DFT $F_d$ through modal fidelity and success probability: 
\begin{equation}
\label{eq:FP}
\sF_W = \frac{\left|\Tr W^\dagger F_d \right|^2}{d^2 \sP_W} \;\;\; ; \;\;\;
\sP_W = \frac{\Tr W^\dagger W}{d}.
\end{equation}

\begin{figure*}[bt!]
\centering
\includegraphics[width=0.8\textwidth]{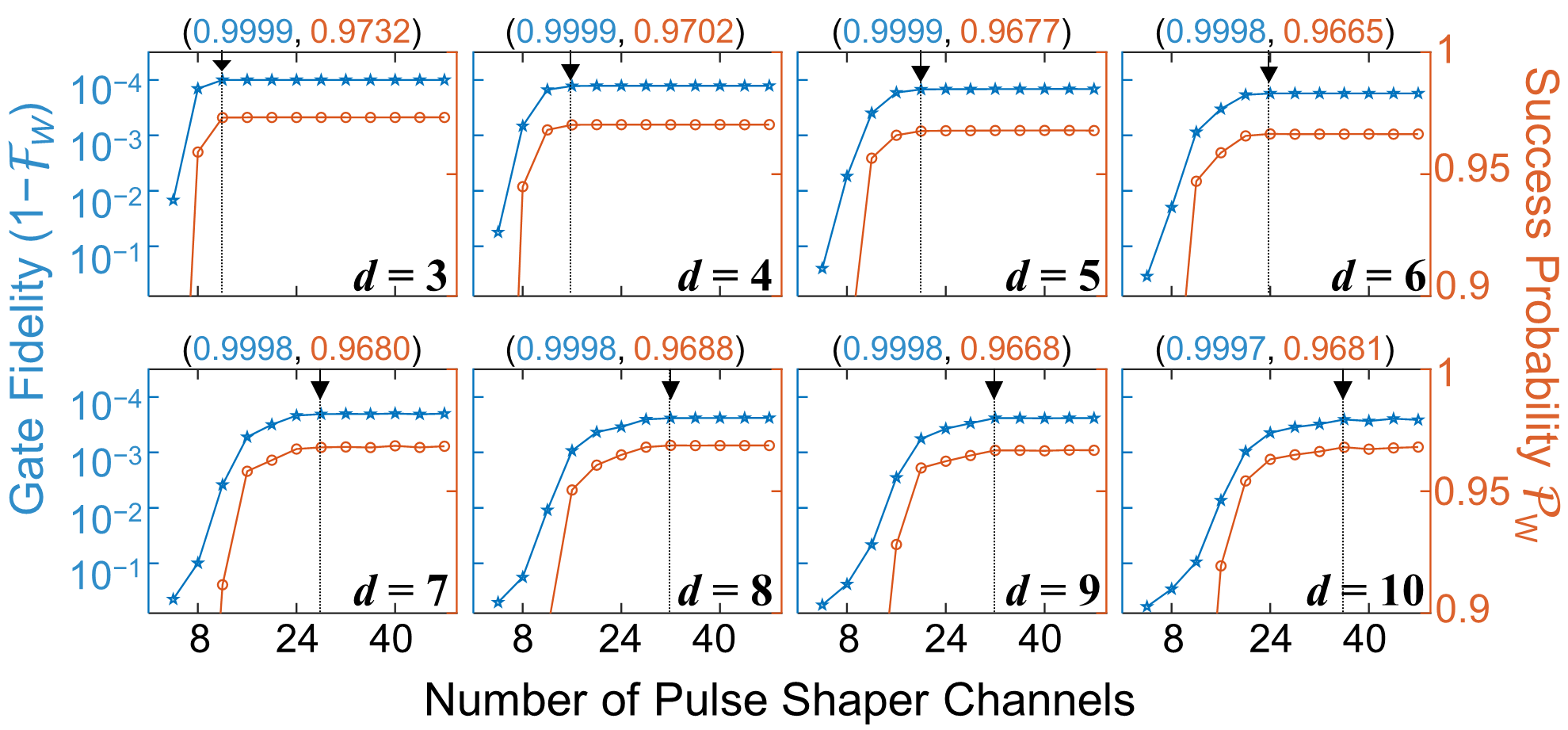}
\caption{DFT solutions for a three-element QFP, where $d-1$ RF harmonics are available for each dimension $d$. The vertical arrows mark the bandwidth required to reach the final cost function value, up to three significant digits; the fidelity and success probability $(\sF_W,\sP_W)$ at these points are provided above each plot.}
\label{BWscale}
\end{figure*}

We truncate the formal infinite-dimensional space of \eqref{eq:W} to $M=64$ modes for numerical simulation, which we have found sufficiently large to eliminate spurious edge effects. Of these $M$ modes, only $B$ are phase-shifted by the pulse shaper (the remaining $M-B$ are passed with zero applied phase), leaving a total of $B+4P$ independent parameters to optimize: $B$ pulse shaper phases $\phi_k$, and amplitude and phase for each harmonic of each EOM. Using particle swarm optimization~\cite{Kennedy1995}, we find the parameter settings which minimize the cost function $C=\sP_W \log_{10}(1-\sF_W)$ as a convenient means to optimize both $\sF_W$ and $\sP_W$ while penalizing the former more strongly~\cite{Pizzimenti2021}. We repeat such optimization tasks for $B\in\{4,8,...,52\}$  to investigate the required number of pulse shaper channels as well. 

As observed in Ref.~\cite{Lu2018a}, a lone single-pass EOM cannot mix $d$ frequency bins equally without at least $\frac{d-1}{2d-1}$ of the input energy scattering into modes outside of the $d$-dimensional subspace. However, by cascading multiple EOMs separated by pulse shapers---the QFP---this scattering can be compensated and high-probability, high-fidelity mixing is possible.
In the first QFP realization of the $d=2$ ($d=3$) DFT, a solution with fidelity $\sF_W=0.9999$ ($\sF_W=0.9999$) and success probability $\sP_W=0.9760$ ($\sP_W=0.9733$) was demonstrated~\cite{Lu2018a}. Interestingly, while the $d=2$ case utilized single-frequency sinewave modulation, the $d=3$ solution required modulation containing both the first and second harmonic. To see if this ``add-RF-harmonic'' rule represents a trend for DFT gates, here we perform additional design simulations to synthesize $d$-dimensional DFT gates on a three-element QFP, in which we consider $d-1$ RF tones in the optimization procedure. The resulting fidelities and success probabilities for $d\leq 10$ and channel numbers $B$ appear in Fig.~\ref{BWscale}. For all $d$, solutions with $\sF_W>0.9997$ and $\sP_W>0.965$ are possible with these resources. 


While the number of elements required for these results is constant (fixed at three), 
the effective number of modes utilized does increase with $d$, as expressed by the vertical lines in Fig.~\ref{BWscale}, which mark where the third significant digit of the cost $\sP_W\log_{10}(1-\sF_W)$ has converged to its limiting value, specifically at $B=12$ ($d=3$), 16 ($d=4$), 20 ($d=5$), 24 ($d=6$), 28 ($d=7$), 32 ($d=8$), 32 ($d=9$), and 36 ($d=10$). The ordered pair $(\sF_W,\sP_W)$ above each plot shows the specific fidelity and success probability for the solution at this value of $B$, which we hereafter refer to as ``minimum bandwidth'' DFT solutions. 
This bandwidth scaling with dimension is consistent with previous observations of a tradeoff between QFP depth and optical bandwidth~\cite{Lukens2017}. From a practical side, accessing additional bandwidth is frequently preferred to adding components---in terms of cost and loss---so that this fixed-depth DFT design procedure appears quite useful. Incidentally, it also seems that the ``$d-1$ RF tone'' rule is unique and well defined: for all simulations we have completed, access to either fewer or more RF harmonics leads, respectively, to noticeable reductions or negligible improvements in DFT gate performance.

\begin{figure*}[bt!]
\centering
\includegraphics[width=0.8\textwidth]{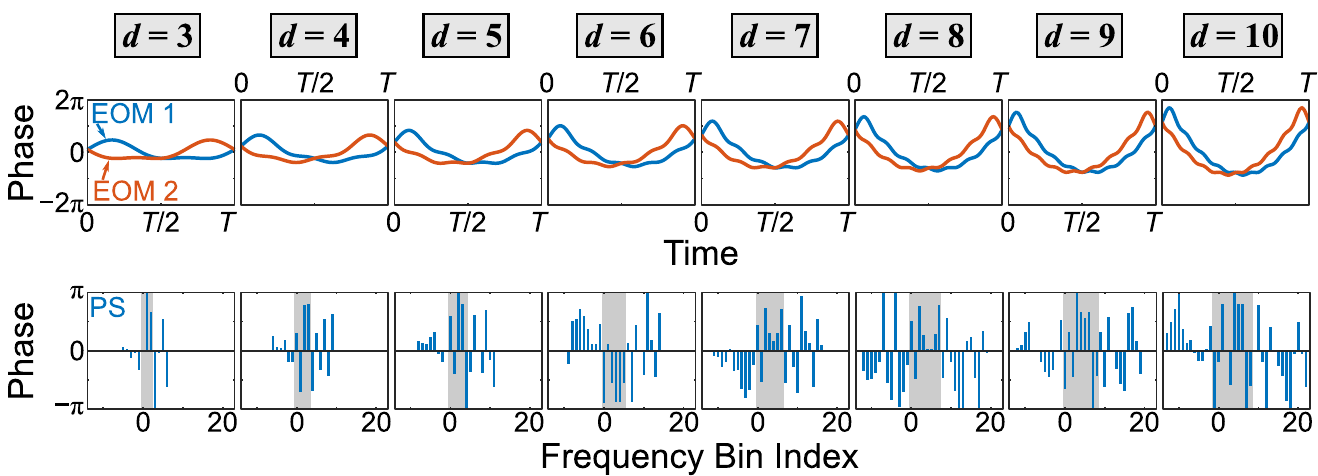}
\caption{Full QFP solutions for the $d$-point DFT gates with bandwidths indicated by the arrows in Fig.~\ref{BWscale}. The top row plots phase modulation patterns for both EOMs over a single temporal period $T=2\pi/\Delta\omega$. The bottom row shows the phases applied to each frequency bin by the central pulse shaper (PS), where gray shading encloses the computational space from bin 0 to $d-1$.}
\label{QFPsol}
\end{figure*}

Figure~\ref{QFPsol} plots the specific EOM modulation patterns [$A(t)$ and $B(t)$ in Eq.~\eqref{eq:c}] and pulse shaper phases [$\phi_k$ in Eq.~\eqref{eq:W}] for the minimum bandwidth solutions designated by arrows in Fig.~\ref{BWscale}. The pulse shaper phase shifts (bottom row) display no obvious trends in their spectrum as $d$ increases. In contrast, the temporal phases on the EOMs (top row) possess a clear single-peak structure that becomes sharper and grows in amplitude with $d$. Moreover, the second EOM pattern is the time-reversed version of the first [$B(t)=A(-t)$] for all $d$, which is especially interesting in that these modulation functions do not cancel each other out for the case of zero applied pulse shaper phase ($\phi_k=0$). 
Our attempts to understand this behavior intuitively have proven unsuccessful, although we suspect a useful explanation should be possible. Nevertheless, the practical value of these recipes for frequency-bin DFTs remains clear, particularly toward on-chip integration where tighter mode spacings could make the synthesis of high-order RF harmonics more manageable.

\textit{Experiment.---}As an application of the DFT for state characterization, we experimentally implement the $d=3$ solution and apply it to a biphoton frequency comb (BFC). While parallel $d=2$ DFTs have been realized~\cite{Lu2018b}, as well as a single-photon $d=3$ DFT~\cite{Lu2018a}, this is the first example combining the two: i.e., parallel frequency-bin DFTs on frequency-bin qutrits. Our experimental design for $d=3$ resembles the scheme discussed in Fig.~\ref{GenScheme}, where we enlist the PPLN biphoton source and use  
a pulse shaper to carve a total of three pairs of frequency-correlated, 20~GHz-spaced, $\sim$10~GHz-wide bins; the bin spacing $\Delta\omega/2\pi=20$~GHz facilitates line-by-line shaping, and RF tones at 20 and 40~GHz are required at each of the QFP's two EOMs. 
As experimental simplifications due to available equipment, the signal and idler photons are transmitted in the same optical fiber and then modulated by a \emph{single} QFP programmed with two parallel DFT gates (separated by a 200~GHz guardband), and coincidences are registered by raster scanning signal and idler WSS filters so that only two SNSPDs are required (rather than the ideal of $2d=6$).

\begin{figure}[b!]
\centering
\includegraphics[width=3.3in]{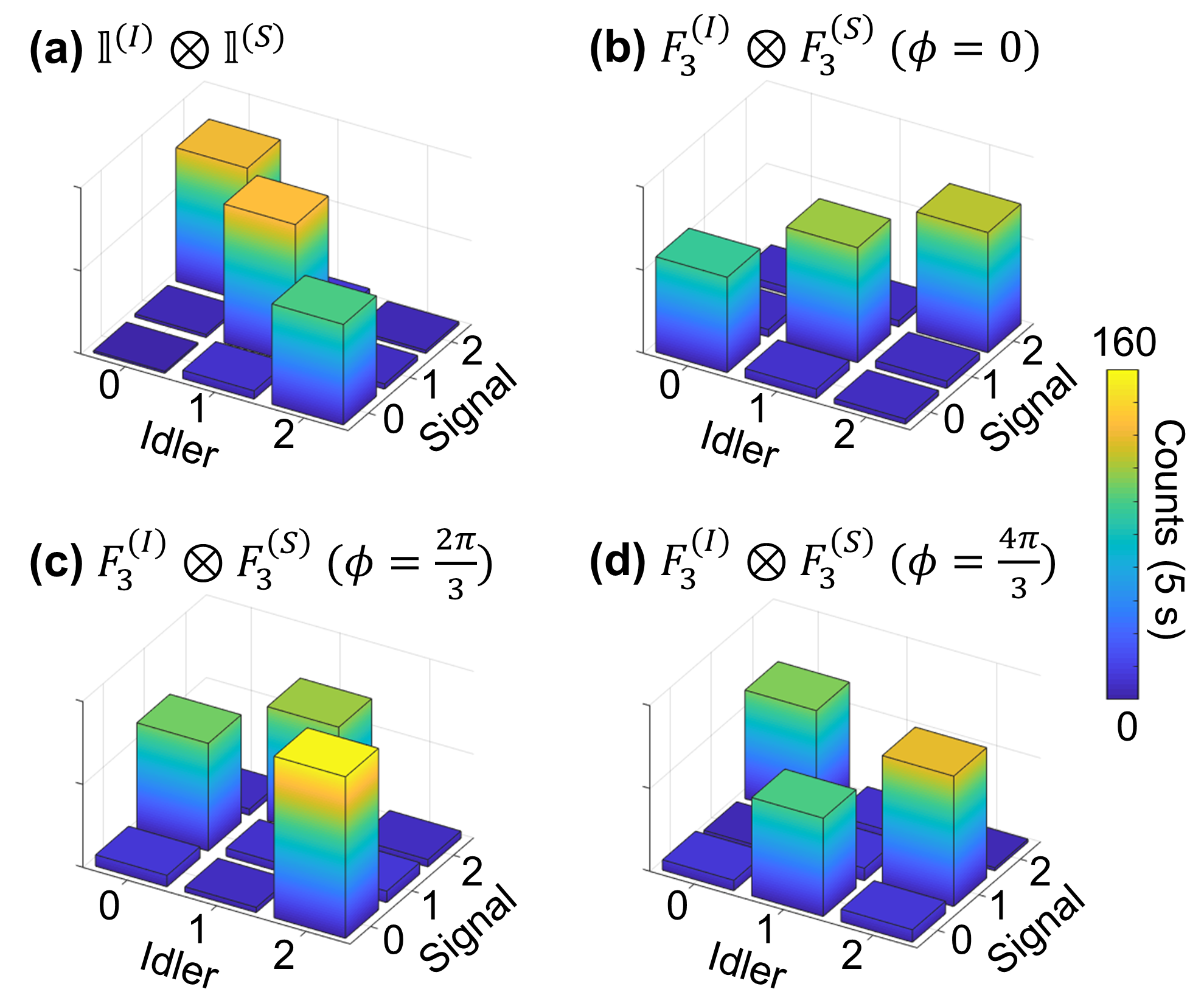}
\caption{Spectrally resolved coincidences after parallel QFP operations. Results correspond to measurements in the logical (a) and DFT bases (b--d) . The latter vary with the phase of the prepared input superposition state $\ket{\phi}$.}
\label{fig3}
\end{figure}

A logical basis measurement (EOMs off) of our $3\times 3$ BFC appears in Fig.~\ref{fig3}(a). Using the front-end pulse shaper to produce biphoton states ideally of the form $\ket{\phi}\propto \ket{02}_{IS} + e^{i\phi}\ket{11}_{IS} + e^{2i\phi}\ket{20}_{IS}$, the measured output coincidences for $\phi\in\{0,2\pi/3,4\pi/3\}$ after parallel DFTs follow in Fig.~\ref{fig3}(b--d): as expected, the results are strongly correlated, with each setting of $\phi$ determining which three pairs of frequency bins are populated. Despite the small number of measurements considered, the observed correlations are sufficient for meaningful inference of the underlying states. Since the prepared states differ only in phase, we can take the logical basis results in (a) as applying to any of the three $\phi$ cases, giving us two sets of nine-outcome measurements for each $\phi$ value, $\mathbbm{1}^{(I)}\otimes \mathbbm{1}^{(S)}$ and $F_3^{(I)}\otimes F_3^{(S)}$.

One useful metric for a bipartite state is the distillable entanglement $E_D$
~\cite{Bennett1996}. While extremely difficult to determine directly, bounds can be obtained from computable quantities. For example, a lower bound can be established from conditional entropies~\cite{Coles2017}, namely: $E_D \geq \log_2 3 - \sH(\mathbbm{1}^{(I)}|\mathbbm{1}^{(S)}) - \sH(F_3^{(I)}|F_3^{(S)})$. Following previous work~\cite{Lu2018b}, we can estimate these entropies directly from the raw counts in Fig.~\ref{fig3}, by positing some unknown nine-element probability distribution $\mathbf{p}=(p_{00}, p_{01},...,p_{22})$ for each panel and sampling the Bayesian posterior distribution formed by a flat Dirichlet prior on $\mathbf{p}$ and a multinomial likelihood for the observed counts. Doing so, we obtain
the $\min E_D$ values in Table~\ref{tab:char}.

In addition, $E_D$ can be upper bounded by the log-negativity $E_\mathcal{N}$~\cite{Vidal2002}, which requires the full density matrix for computation. Utilizing Bayesian quantum state tomography~\cite{Blume2010, Lukens2020b}, which returns uncertainties commensurate with the data gathered, we can indeed estimate the full quantum state with these results, irrespective of their informational completeness. 
Applying the specific Bayesian workflow described in Ref.~\cite{Lu2021}---which employs a Bures prior distribution and accounts for raster scanning and $\sP_W<1$ through a Poissonian likelihood---we find the fidelities for each state as shown in Table~\ref{tab:char}, defined as $\sF_\rho = \braket{\phi|\rho|\phi}$. 
Since the outcomes are strongly correlated, the inferred states have relatively small uncertainty, even with measurements in two bases only. Computing log-negativity $E_\mathcal{N}$, we obtain a complete interval for $E_D$ of approximately $E_D \in [0.4,1.3]$ ebits for the states considered. This range is quite wide; we suspect that the much higher values for $E_\mathcal{N}$ result from the fact it applies quantum state constraints, in contrast to the entropic bound which treats the measurement results as raw probabilities. In other words, quantum state tomography is based on the assumption of a single ground truth state behind all measurement sets involved, as well as a physical model (Born's rule) connecting this state to the observed outcomes, whereas our entropic calculation views the detection scenarios as isolated probability distributions with no direct connection to quantum measurement theory. It would be interesting to explore how the $E_D$ range may narrow with higher-fidelity results, which are limited here primarily by the resolution of the state preparation and measurement pulse shapers.

Finally, as an aside, we note that one can alternatively view the $\phi$ and $2\phi$ phase shifts imparted by the front-end pulse shaper as part of the measurement process, rather than state preparation, and pool all results of Fig.~\ref{fig3} into a single likelihood to estimate the state \emph{before} the pulse shaper (ideally $\ket{\phi=0}$). Doing so, we find $\sF_\rho=0.81\pm 0.02$ and $E_\mathcal{N}=1.32\pm0.03$, comparable to the values in Table~\ref{tab:char}, albeit with a slight increase in entanglement likely resulting from access to correlations in four---rather than just two---bases in the inference process.

\begin{table}[t!]
    \centering
    \begin{tabular}{|c|c|c|c|}
    \hline
         $\phi$ & $\min E_D$ [ebits] & $\sF_\rho$ & $E_\mathcal{N}$ [ebits]  \\ \hline
         0 & $0.41\pm0.09$ & $0.80\pm0.02$ & $1.30\pm0.03$ \\
         $2\pi/3$ & $0.41\pm0.09$ & $0.80\pm0.02$ & $1.30\pm0.04$ \\
         $4\pi/3$ & $0.36\pm0.09$ & $0.78\pm0.02$ & $1.28\pm0.04$ \\\hline
    \end{tabular}
    \caption{Bayesian inference of Fig.~\ref{fig3} data: distillable entanglement bound $\min E_D$, fidelity $\sF_\rho$, and log-negativity $E_\mathcal{N}$.}
    \label{tab:char}
\end{table}

\textit{Discussion.---}The frequency-bin DFT designs introduced and analyzed here reveal intriguing opportunities for quantum information processing tasks in frequency encoding, including QKD. 
Indeed, the high success probabilities of our DFT solutions actually address a challenge shared by time-bin DFT measurements as well: the satellite pulses produced by passive delay interferometers lead to an effective measurement probability of $1/d$~\cite{Islam2017a, Islam2017b}. The frequency-bin DFTs here with near-unity success 
thus render our design similar in spirit to more complex active time-bin interferometers~\cite{Wang2015a, Wang2018b, Lukens2018a} that can in principle reach unit DFT measurement efficiency. As another application, aligned with the original motivation for the QFP~\cite{Lukens2017}, our DFT is precisely the operation required for a quantum interconnect of frequency-disparate matter qubits. Consider $d$ qubits, each in a tuned lambda energy scheme, such that pump-induced excitation from the ground to excited state is accompanied by a emission of a single photon at frequency $\omega_n=\omega_0 + n\Delta\omega$---a distinct value for each qubit $n$. By coupling these photonic modes into a single bus waveguide terminated in a $d$-point DFT, detection of a single photon in output frequency mode $m$ will herald generation of the $W$-like entangled matter state $\ket{\psi_m} = \frac{1}{\sqrt{d}} \sum_{n=0}^{d-1} e^{2\pi\imag mn/d} \ket{0\cdots 0 1_n 0 \cdots 0}$, a spectral version of the  Duan--Lukin--Cirac--Kimble (DLCZ) interconnect~\cite{Duan2001} generalized to $d>2$ qubits as explored previously in the spatial domain~\cite{Lougovski2009, Choi2010}. And since the frequency-bin version leverages a single spatial mode in a bus waveguide, environmental fluctuations are shared by all interfering bins, minimizing the need for active wavelength-level path stabilization and automatically circumventing an inherent challenge of path-encoded DLCZ-type protocols. 

Finally, the fact that DFT gates of increasing dimension are realizable without the addition of more pulse shapers and EOMs makes our design of particular value with near-term technology. Admittedly, 
at our experimental value $\Delta\omega/2\pi = 20$~GHz, the $d=10$ solution would require the coherent combination of 9 RF tones up to a maximum frequency of 180~GHz---a questionable prospect both in terms of microwave engineering and raw bandwidth. Yet although the minimum spacing $\Delta\omega$ is limited in our case by the resolution of the diffractive pulse shaper, much tighter frequency spacings should be possible with integrated pulse shapers based on MRR add-drop filters~\cite{Agarwal2006, Khan2010, Wang2015b}. For example, at $\Delta\omega/2\pi= 5$~GHz,
the maximum modulation frequency for a 10-point DFT drops to 45~GHz, sufficiently low that even direct digital synthesis of the total waveform should be feasible. 
We therefore envision on-chip QFPs as the most promising route for the high-dimensional frequency-bin mixers discovered here.

\begin{acknowledgments}
Preliminary results were presented at CLEO 2021 as paper number FTu1N.8. We thank AdvR, Inc., for loaning the PPLN ridge waveguide. This research was performed in part at Oak Ridge National Laboratory, managed by UT-Battelle, LLC, for the U.S. Department of Energy under contract no. DE-AC05-00OR22725. Funding was provided by the U.S. Department of Energy, Office of Science, Advanced  Scientific Computing Research, Early Career Research Program (Field Work Proposal ERKJ353), the Air Force Research Laboratory, Small Business Technology Transfer Program (AFRL Prime Order No. FA8750-20-P-1705), and the National Science Foundation (1839191-ECCS, 1747426-DMR).
\end{acknowledgments}


\begin{thebibliography}{10}
\newcommand{\enquote}[1]{``#1''}

\bibitem{Wootters1989}
W.~K. Wootters and B.~D. Fields, \enquote{Optimal state-determination by
  mutually unbiased measurements,} {\em Ann. Phys.}
  \textbf{191}, 363--381 (1989).

\bibitem{Durt2010}
T.~Durt, B.-G. Englert, I.~Bengtsson, and K.~\.{Z}yczkowski, \enquote{On
  mutually unbiased bases,} {\em Int. J. Quant. Inf.}
  \textbf{8}, 535--640 (2010).

\bibitem{Adamson2010}
R.~B.~A. Adamson and A.~M. Steinberg, \enquote{Improving quantum state
  estimation with mutually unbiased bases,} {\em Phys. Rev.
  Lett.} \textbf{105}, 030406 (2010).

\bibitem{Scarani2009}
V.~Scarani, H.~Bechmann-Pasquinucci, N.~J. Cerf, M.~Du\u{s}ek,
  N.~L\"{u}tkenhaus, and M.~Peev, \enquote{The security of practical quantum
  key distribution,} {\em Rev. Mod. Phys.} \textbf{81},
  1301--1350 (2009).

\bibitem{Cerf2002}
N.~J. Cerf, M.~Bourennane, A.~Karlsson, and N.~Gisin, \enquote{Security of
  quantum key distribution using $d$-level systems,}
  {\em Phys. Rev. Lett.} \textbf{88}, 127902 (2002).

\bibitem{Sheridan2010}
L.~Sheridan and V.~Scarani, \enquote{Security proof for quantum key
  distribution using qudit systems,} {\em Phys. Rev. A}
  \textbf{82}, 030301 (2010).

\bibitem{Spengler2012}
C.~Spengler, M.~Huber, S.~Brierley, T.~Adaktylos, and B.~C. Hiesmayr,
  \enquote{Entanglement detection via mutually unbiased bases,}
  {\em Phys. Rev. A} \textbf{86}, 022311 (2012).

\bibitem{Coles2017}
P.~J. Coles, M.~Berta, M.~Tomamichel, and S.~Wehner, \enquote{Entropic
  uncertainty relations and their applications,} {\em Rev.
  Mod. Phys.} \textbf{89}, 015002 (2017).

\bibitem{Hillerkuss2010}
D.~Hillerkuss, M.~Winter, M.~Teschke, A.~Marculescu, J.~Li, G.~Sigurdsson,
  K.~Worms, S.~B. Ezra, N.~Narkiss, W.~Freude, and J.~Leuthold, \enquote{Simple
  all-optical fft scheme enabling tbit/s real-time signal processing,}
  {\em Opt. Express} \textbf{18}, 9324--9340 (2010).

\bibitem{Hillerkuss2011}
D.~Hillerkuss, R.~Schmogrow, T.~Schellinger, M.~Jordan, M.~Winter, G.~Huber,
  T.~Vallaitis, R.~Bonk, P.~Kleinow, F.~Frey, M.~Roeger, S.~Koenig, A.~Ludwig,
  A.~Marculescu, J.~Li, M.~Hoh, M.~Dreschmann, J.~Meyer, M.~Huebner, J.~Becker,
  C.~Koos, W.~Freude, and J.~Leuthold, \enquote{26 {Tbit} s$^{-1}$ line-rate
  super-channel transmission utilizing all-optical fast {Fourier} transform
  processing,} {\em Nat. Photonics} \textbf{5}, 364--371
  (2011).

\bibitem{Islam2017a}
N.~T. Islam, C.~Cahall, A.~Aragoneses, A.~Lezama, J.~Kim, and D.~J. Gauthier,
  \enquote{Robust and stable delay interferometers with application to
  $d$-dimensional time-frequency quantum key distribution,}
  {\em Phys. Rev. Applied} \textbf{7}, 044010 (2017).

\bibitem{Islam2017b}
N.~T. Islam, C.~C.~W. Lim, C.~Cahall, J.~Kim, and D.~J. Gauthier,
  \enquote{Provably secure and high-rate quantum key distribution with time-bin
  qudits,} {\em Sci. Adv.} \textbf{3}, e1701491 (2017).

\bibitem{Wang2018a}
J.~Wang, S.~Paesani, Y.~Ding, R.~Santagati, P.~Skrzypczyk, A.~Salavrakos,
  J.~Tura, R.~Augusiak, L.~Man{\v c}inska, D.~Bacco, D.~Bonneau, J.~W.
  Silverstone, Q.~Gong, A.~Ac{\'\i}n, K.~Rottwitt, L.~K. Oxenl{\o}we, J.~L.
  O{\textquoteright}Brien, A.~Laing, and M.~G. Thompson,
  \enquote{Multidimensional quantum entanglement with large-scale integrated
  optics,} {\em Science} \textbf{360}, 285--291 (2018).

\bibitem{Spagnolo2013}
N.~Spagnolo, C.~Vitelli, L.~Aparo, P.~Mataloni, F.~Sciarrino, A.~Crespi,
  R.~Ramponi, and R.~Osellame, \enquote{Three-photon bosonic coalescence in an
  integrated tritter,} {\em Nat. Commun.} \textbf{4}, 1606
  (2013).

\bibitem{Giovannini2013}
D.~Giovannini, J.~Romero, J.~Leach, A.~Dudley, A.~Forbes, and M.~J. Padgett,
  \enquote{Characterization of high-dimensional entangled systems via mutually
  unbiased measurements,} {\em Phys. Rev. Lett.}
  \textbf{110}, 143601 (2013).

\bibitem{Bavaresco2018}
J.~Bavaresco, N.~H. Valencia, C.~Kl{\"o}ckl, M.~Pivoluska, P.~Erker, N.~Friis,
  M.~Malik, and M.~Huber, \enquote{Measurements in two bases are sufficient for
  certifying high-dimensional entanglement,} {\em Nat.
  Phys.} \textbf{14}, 1032--1037 (2018).

\bibitem{Ecker2019}
S.~Ecker, F.~Bouchard, L.~Bulla, F.~Brandt, O.~Kohout, F.~Steinlechner,
  R.~Fickler, M.~Malik, Y.~Guryanova, R.~Ursin, and M.~Huber,
  \enquote{Overcoming noise in entanglement distribution,}
  {\em Phys. Rev. X} \textbf{9}, 041042 (2019).

\bibitem{Lukens2017}
J.~M. Lukens and P.~Lougovski, \enquote{Frequency-encoded photonic qubits for
  scalable quantum information processing,} {\em Optica}
  \textbf{4}, 8--16 (2017).

\bibitem{Lu2018a}
H.-H. Lu, J.~M. Lukens, N.~A. Peters, O.~D. Odele, D.~E. Leaird, A.~M. Weiner,
  and P.~Lougovski, \enquote{Electro-optic frequency beam splitters and
  tritters for high-fidelity photonic quantum information processing,}
  {\em Phys. Rev. Lett.} \textbf{120}, 030502 (2018).

\bibitem{Lukens2020a}
J.~M. Lukens, H.-H. Lu, B.~Qi, P.~Lougovski, A.~M. Weiner, and B.~P. Williams,
  \enquote{All-optical frequency processor for networking applications,}
  {\em J. Lightwave Technol.} \textbf{38}, 1678--1687
  (2020).

\bibitem{Kennedy1995}
J.~Kennedy and R.~Eberhart, \enquote{Particle swarm optimization,} in
  \emph{Proc. Int. Conf. Neural Netw.}, , vol.~4 (1995), pp. 1942--1948.

\bibitem{Pizzimenti2021}
A.~J. Pizzimenti, J.~M. Lukens, H.-H. Lu, N.~A. Peters, S.~Guha, and C.~N.
  Gagatsos, \enquote{Non-gaussian photonic state engineering with the quantum
  frequency processor,} {\em Phys. Rev. A} \textbf{104},
  062437 (2021).

\bibitem{Lu2018b}
H.-H. Lu, J.~M. Lukens, N.~A. Peters, B.~P. Williams, A.~M. Weiner, and
  P.~Lougovski, \enquote{Quantum interference and correlation control of
  frequency-bin qubits,} {\em Optica} \textbf{5}, 1455--1460
  (2018).

\bibitem{Bennett1996}
C.~H. Bennett, D.~P. DiVincenzo, J.~A. Smolin, and W.~K. Wootters,
  \enquote{Mixed-state entanglement and quantum error correction,}
  {\em Phys. Rev. A} \textbf{54}, 3824--3851 (1996).

\bibitem{Vidal2002}
G.~Vidal and R.~F. Werner, \enquote{Computable measure of entanglement,}
  {\em Phys. Rev. A} \textbf{65}, 032314 (2002).

\bibitem{Blume2010}
R.~Blume-Kohout, \enquote{Optimal, reliable estimation of quantum states,}
  {\em New J. Phys.} \textbf{12}, 043034 (2010).

\bibitem{Lukens2020b}
J.~M. Lukens, K.~J.~H. Law, A.~Jasra, and P.~Lougovski, \enquote{A practical
  and efficient approach for {Bayesian} quantum state estimation,}
  {\em New J. Phys.} \textbf{22}, 063038 (2020).

\bibitem{Lu2021}
H.-H. Lu, K.~V. Myilswamy, R.~S. Bennink, S.~Seshadri, M.~S. Alshaykh, J.~Liu,
  T.~J. Kippenberg, D.~E. Leaird, A.~M. Weiner, and J.~M. Lukens, \enquote{Full
  quantum state tomography of high-dimensional on-chip biphoton frequency combs
  with randomized measurements,} {\em arXiv:2108.04124}
  (2021).

\bibitem{Wang2015a}
S.~Wang, Z.-Q. Yin, W.~Chen, D.-Y. He, X.-T. Song, H.-W. Li, L.-J. Zhang,
  Z.~Zhou, G.-C. Guo, and Z.-F. Han, \enquote{Experimental demonstration of a
  quantum key distribution without signal disturbance monitoring,}
  {\em Nat. Photonics} \textbf{9}, 832--836 (2015).

\bibitem{Wang2018b}
S.~Wang, Z.-Q. Yin, H.~F. Chau, W.~Chen, C.~Wang, G.-C. Guo, and Z.-F. Han,
  \enquote{Proof-of-principle experimental realization of a qubit-like
  qudit-based quantum key distribution scheme,} {\em Quantum
  Sci. Technol.} \textbf{3}, 025006 (2018).

\bibitem{Lukens2018a}
J.~M. Lukens, N.~T. Islam, C.~C.~W. Lim, and D.~J. Gauthier,
  \enquote{Reconfigurable generation and measurement of mutually unbiased bases
  for time-bin qudits,} {\em Appl. Phys. Lett.}
  \textbf{112}, 111102 (2018).

\bibitem{Duan2001}
L.-M. Duan, M.~D. Lukin, J.~I. Cirac, and P.~Zoller, \enquote{Long-distance
  quantum communication with atomic ensembles and linear optics,}
  {\em Nature} \textbf{414}, 413--418 (2001).

\bibitem{Lougovski2009}
P.~Lougovski, S.~J. van Enk, K.~S. Choi, S.~B. Papp, H.~Deng, and H.~J. Kimble,
  \enquote{Verifying multipartite mode entanglement of {W} states,}
  {\em New J. Phys.} \textbf{11}, 063029 (2009).

\bibitem{Choi2010}
K.~S. Choi, A.~Goban, S.~B. Papp, S.~J. Van~Enk, and H.~J. Kimble,
  \enquote{Entanglement of spin waves among four quantum memories,}
  {\em Nature} \textbf{468}, 412--416 (2010).

\bibitem{Agarwal2006}
A.~{Agarwal}, P.~{Toliver}, R.~{Menendez}, S.~{Etemad}, J.~{Jackel},
  J.~{Young}, T.~{Banwell}, B.~E. {Little}, S.~T. {Chu}, {Wei Chen}, {Wenlu
  Chen}, J.~{Hryniewicz}, F.~{Johnson}, D.~{Gill}, O.~{King}, R.~{Davidson},
  K.~{Donovan}, and P.~J. {Delfyett}, \enquote{Fully programmable
  ring-resonator-based integrated photonic circuit for phase coherent
  applications,} {\em J. Lightwave Technol.} \textbf{24},
  77--87 (2006).

\bibitem{Khan2010}
M.~H. Khan, H.~Shen, Y.~Xuan, L.~Zhao, S.~Xiao, D.~E. Leaird, A.~M. Weiner, and
  M.~Qi, \enquote{Ultrabroad-bandwidth arbitrary radiofrequency waveform
  generation with a silicon photonic chip-based spectral shaper,}
  {\em Nat. Photonics} \textbf{4}, 117--122 (2010).

\bibitem{Wang2015b}
J.~Wang, H.~Shen, L.~Fan, R.~Wu, B.~Niu, L.~T. Varghese, Y.~Xuan, D.~E. Leaird,
  X.~Wang, F.~Gan, A.~M. Weiner, and M.~Qi, \enquote{Reconfigurable
  radio-frequency arbitrary waveforms synthesized in a silicon photonic chip,}
  {\em Nat. Commun.} \textbf{6}, 5957 (2015).

\end{thebibliography}
\end{document}